%Paper: 9112066
%From: KONSTANT@SUHEP.PHY.SYR.EDU
%Date: Fri, 20 Dec 1991 22:29:46 EST

\input harvmac
\noblackbox
%\draftmode

\def\l{\lambda}

\def\meas{\oint {dz\over 2 \pi i z}}
\def\part{\partial}
\def\lb{\lbrack}
\def\rb{\rbrack}
\def\nl{ {N\over\l}  }
\def\nl2{ {N\over {2\l}}  }
\def\zd{ z\partial_z }

\def\None{ N^{-{1\over {2k+1} } } }
\def\Ntwo{ N^{-{2\over {2k+1} } } }

\def\cp{ {\cal P } }
\def\cq{ {\cal Q } }
\def\cqt{ {\tilde {\cal Q }} }
\def\cpt{ {\tilde {\cal P }} }
\def\bp{ {\bf P} }
\def\bpt{ {\bf {\tilde P} } }

\def\bpd{ {\bf P}^{\dagger } }
\def\R{ {\rm R} }
\def\d{\delta}
\def\t{\tau}
\def\to{\tau_1}
\def\tt{\tau_2}
\def\pa{\partial}
\def\o{\over}
\def\Gr{Gr^{(0)}}
\def\vo{V_1}
\def\vt{V_2}

\def\so{S_1}
\def\st{S_2}
\def\sod{S_1^{\dagger}}
\def\std{S_2^{\dagger}}
\def\soi{S_1^{-1}}
\def\sti{S_2^{-1}}
\def\ut{{\tilde u}}
\def\a{\alpha}
\def\ps{\psi}
\def\psd{\psi^{\dagger}}
\def\ph{\phi}
\def\C{ {\rm C} }
\def\L{ {\rm L} }
\def\B{ {\rm B} }
\def\F{ {\rm F} }
\def\GL{ { {\rm GL}(\infty) } }
\def\gl{ { {\rm gl}(\infty) } }
\def\Z{ {\bf{\rm Z}} }
\def\poo{P_{00}(z)}
\def\pio{P_{10}(z)}
\def\poi{P_{01}(z)}
\def\pii{P_{11}(z)}
%\baselineskip=24 true pt
%\hsize=6 true in \hoffset=.25 true in
%\vsize=8.5true in \voffset = .2 true in

%-----------------------title-------------------------
\Title{\vbox{\baselineskip12pt\hbox{SU-4238-497}\hbox{NSF-ITP-91-133}}}
{{\vbox{\centerline{The Solution Space of the Unitary Matrix}
\vskip2pt \centerline{Model String Equation and the }
\vskip2pt \centerline{Sato Grassmannian }
}}}
\bigskip
\centerline{ Konstantinos N. Anagnostopoulos\footnote{$^1$}
{ E-mail: Konstant@suhep.bitnet; Bowick@suhep.bitnet.}
{\rm and}  $~$Mark J. Bowick$^1$}
\medskip
\centerline{ Physics Department }
\centerline{ Syracuse University}
\centerline{ Syracuse, NY 13244-1130, USA}
\bigskip
\centerline{ Albert Schwarz\footnote{$^2$}
{ E-mail: Asschwarz@ucdavis.edu}}
\centerline{Department of Mathematics}
\centerline{ University of California}
\centerline{ Davis, CA 95616, USA}
\bigskip
\centerline{\bf Abstract}
\medskip
The space of all solutions to the string equation of the symmetric unitary
one-matrix model is determined. It is shown that the string equation is
equivalent to simple conditions on points $V_1$ and $V_2$ in the big cell $\Gr$
of the Sato Grassmannian $Gr$. This is a consequence of a well-defined
continuum limit in which the
string equation has the simple form $\lb \cp ,\cq_- \rb =\hbox{\rm 1}$,
with $\cp$
and $\cq_-$ $2\times 2$ matrices of differential operators. These
conditions on $V_1$ and $V_2$ yield a simple system of first order differential
equations whose analysis determines the space of all solutions to the string
equation. This geometric formulation leads directly to the Virasoro
constraints $\L_n\,(n\geq 0)$, where $\L_n$ annihilate the two modified-KdV
$\t$-functions whose product gives the partition function of the Unitary Matrix
Model.
\noindent
\Date{December 20, 1991}
\vfill \eject
%
%
%--------------------------------references----------------------
%-------------------------------text------------------------
%
\centerline{\bf 1. Introduction} \bigskip
Matrix models form a rich class of quantum statistical mechanical systems
defined by partition functions of the form
$\int \,dM\,{\rm e}^{-{N\o \l} \tr V(M)}$, where $M$ is an
$N\times N$ matrix and the
Hamiltonian $\tr V(M)$ is some well defined function of $M$. They were
originally introduced to study complicated systems, such as heavy nuclei,
in which the quantum mechanical Hamiltonian had to be considered random within
some universality class \nref\meht{Mehta, M. L.:
Random Matrices.  ${\rm 2}^{\rm nd}$ Edition, Boston: Academic Press 1991
}\nref\dyso{
Dyson, F. J.:
Statistical Theory of the Energy Levels of Complex Systems I.
J. Math. Phys. {\bf 3} 140-156 (1962)}\nref\dyst{
Dyson, F. J.:
Statistical Theory of the Energy Levels of Complex Systems II.
J. Math. Phys.
{\bf 3}  157-165 (1962)}\nref\dysth{
Dyson, F. J.:
Statistical Theory of the Energy Levels of Complex Systems III.
J. Math. Phys.{\bf 3}  166-175 (1962)}
\refs{\meht , \dysth} .

Unitary Matrix Models (UMM), in which $M$ is a unitary matrix $U$, form a
particularly rich class of matrix models. When $V(U)$ is self adjoint we will
call the model symmetric. The simplest case is given by
$V(U)=U+U^{\dagger}$ and describes two dimensional quantum
chromodynamics \nref\hft{ 't Hooft, G.: A Planar Diagram Theory for Strong
Interactions. Nucl. Phys.
{\bf B72}, 461-473 (1974)}\nref\gros{ Gross, D. and  Witten, E.: Possible
Third-Order Phase Transition in the Large-N Lattice Gauge Theory.
 Phys. Rev. {\bf D21},  446-453 (1980)}\nref\wadia{Wadia, S.~R.:
Dyson-Schwinger
Equations Approach to the Large-N Limit: Model Systems and String
Representation of Yang-Mills Theory.
Phys. Rev. {\bf D24},  970-978 (1981)}\refs{\hft{--}\wadia}
with gauge
group $U(N)$. The partition function of this theory can be evaluated in the
large-$N$ (planar) limit in which $N$ is taken to infinity with
$\l=g^2N$ held fixed, where $g$ is the gauge coupling.
The theory has a third order
phase transition at $\l_c=2$ \gros .
Below $\l_c$ the eigenvalues ${\rm e}^{i\a_j}$ of $U$
lie within a finite domain about $\a=0$ of the form
$\lb -\a_c,\a_c \rb$ with $\a_c <\pi$. The size of this domain increases as
$\l$ increases until the eigenvalues range over the entire circle at $\l=\l_c$.

In the last two years, matrix models have received extensive attention as
discrete models of two dimensional gravity. In this context, the one-matrix
Hermitian
Matrix Models (HMM), in which $M$ is a Hermitian matrix, are the clearest to
interpret since a given cellular decomposition of a two dimensional surface
is dual to a Feynman diagram of a zero dimensional quantum field theory with
action $\tr V(M)$. In the double scaling
limit of these models, the potential can be
tuned to a one parameter family of multicritical points labelled by an integer
$m$. This scaling limit is defined by
$N$ going to infinity and $\l\rightarrow\l_c$
with $t=(1-{n\o N})N^{2m\o 2m+1}$ and
$y=(1-{\l\o \l_c})N^{2m\o 2m+1}$ held fixed. This requires simultaneously
adjusting m couplings in the
potential to their critical values. At these multicritical points the
entire partition function (including the sum over topologies) is given by a
single differential equation (the ``string equation") and can serve as a
non-perturbative definition of two dimensional gravity coupled to conformal
matter \nref\brka{ Br\'ezin, E. and  Kazakov, V.:
Exactly Solvable Field Theories of Closed Strings. Phys. Lett. {\bf B236},
144-149 (1990)}\nref\dosh{Douglas, M. and  Shenker, S.:
Strings in Less Than One Dimension.  Nucl. Phys. {\bf B335},  635-654
(1990)}\nref\grmi{ Gross, D. and  Migdal, A.:
Nonperturbative Two-Dimensional Quantum Gravity. Phys. Rev. Lett. {\bf 64},
 127-130 (1990).}\nref\grmit{ Gross, D. and  Migdal, A.:
A Nonperturbative Treatment of Two-Dimensional Quantum Gravity. Nucl. Phys.
 {\bf B340},  333-365 (1990)}\refs{\brka{--}\grmit}.
This multicriticality may also be described by universal cross-over
behaviour in the tail of the distribution of the eigenvalues \nref\brbo{
Bowick, M. J.  and Br\'ezin, E.:
Universal Scaling of the Tail of the Density of Eigenvalues in Random Matrix
Models. Phys. Lett.
{\bf B268},  21-28 (1991)}\brbo .

UMM have also been solved in the double scaling limit \nref\peri{
Periwal, V.  and Shevitz, D.:
Unitary Matrix Models as Exactly Solvable String Theories.
Phys. Rev. Lett. {\bf 64},  1326-1329 (1990)}\nref\shev{Periwal, V.  and
Shevitz, D.:
Exactly Solvable Unitary Matrix Models: Multicritical Potentials and
Correlations. Nucl. Phys. {\bf B344},  731-746 (1990)}\nref\neub{
Neuberger, H.:
Scaling Regime at the Large-N Phase Transition of Two-Dimensional Pure Gauge
Theories. Nucl. Phys.
{\bf B340},  703-720 (1990)}\nref\neubt{
Neuberger, H.:
Non Perturbative Contributions In Models With A Non-Analytic Behavior at
Infinite N. Nucl. Phys. {\bf B179},  253-282 (1981)}\nref\deme{
Demeterfi, K.
and Tan, C. I.:
String Equations From Unitary Matrix Models. Mod. Phys. Lett.
 {\bf A5},  1563-1574 (1990)}\refs{\peri{--}\deme}
and
their general features are very similar to the HMM. At finite $N$ they exhibit
integrable flows in the parameters of the potential similar
to the HMM \nref\mart{Martinec, E.: On the Origin of Integrability in Matrix
Models. Commun. Math. Phys.
{\bf 138}, 437-449 (1991)}\nref\bowi{Bowick, M.~J., Morozov, A. and
Shevitz, D.: Reduced Unitary Matrix Models and the Hierarchy of $\t$-Functions.
Nucl. Phys. {\bf B354},  496-530 (1991)}\nref\crnka
{Crnkovi\'c, \v C., Douglas, M.  and Moore, G.:
Physical Solutions for Unitary Matrix Models. Nucl. Phys. {\bf B360},
 507-523 (1991)}\nref\crnkb
{Crnkovi\'c, \v C.  and Moore, G.:
Multicritical Multi-Cut Matrix Models.
Phys. Lett. {\bf B257},  322-328 (1991)}\refs{\mart{--}\crnkb}
and in
the double scaling limit they lie in the same universality class as the
double-cut HMM \nref\cdm{Crnkovi\'c, \v C., Douglas, M.  and Moore, G.:
Loop Equations and the Topological Structure of Multi-Cut Models.
Preprint YCTP-P25-91 and RU-91-36}\nref\holo{
Hollowood, T., Miramontes, L., Pasquinucci, A.  and Nappi, C.:
Hermitian vs. Anti-Hermitian 1-Matrix Models and their Hierarchies.
Preprint IASSNS-HEP-91/59 and PUPT-1280}\refs{\crnka{--}\holo}.
The world sheet interpretation
of the UMM is not, however, very clear \cdm .
In view of this it seems worthwhile to explore their structure further.

It is well known \nref\doug{Douglas, M. R.:
Strings in Less Than One Dimensions and the Generalized KdV Hierarchies.
Phys. Lett. {\bf B238},  176-180 (1990)}\doug\
that
the string equation of the $(p,q)$ HMM can be described as an operator equation
$\lb P,Q\rb=1$, where $P$ and $Q$ are scalar ordinary
differential operators of order
$p$ and $q$ respectively. They are the well defined scaling limits of the
operators of multiplication and differentiation by the eigenvalues of the
HMM on the orthonormal polynomials used to
solve the model. The set of solutions to the string equation
$\lb P,Q\rb=1$ was analyzed in \nref\aschw{
Schwarz, A.:
On Solutions to the String Equation.
Mod. Phys. Lett. {\bf A6},  2713-2725 (1991)}\aschw\
by means of the Sato Grassmannian $Gr$. It was proved that every solution of
the string equation corresponds to a point in the big cell $\Gr$ of $Gr$
satisfying certain conditions. This fact was used to give a derivation of
the Virasoro and $W$-constraints obtained in \nref\dvv{
Dijkgraaf, R., Verlinde,  H.  and Verlinde, E.:
Loop Equations and Virasoro Constraints in Non-Perturbative Two-Dimensional
Quantum Gravity. Nucl. Phys.
{\bf B348},  435-456 (1991)}\nref\fknc{Fukuma, M.
Kawai, H.  and Nakayama, R.:
Continuum Schwinger-Dyson Equations and Universal Structures in Two-Dimensional
Quantum Gravity.
Int. J. Mod. Phys.
{\bf A6},  1385-1406 (1991)}\refs{\dvv , \fknc}
along the lines of \nref\kasch{
Kac, V.  and Schwarz, A.:
Geometric Interpretation of the Partition Function of 2D Gravity.
Phys. Lett. {\bf B257}, 329-334 (1991)}\nref\fknb{
Fukuma, M., Kawai,  H.  and Nakayama, R.: Infinite-Dimensional
Grassmannian Structure of Two-Dimensional Gravity.
Preprint UT-572-TOKYO}\nref\schwa{Schwarz, A.:
On Some Mathematical Problems of 2D-Gravity and $W_h$-Gravity.
Mod. Phys. Lett. {\bf A6},  611-616 (1991)}\nref\fkna{
Fukuma, M., Kawai,  H. and Nakayama, R.: Explicit Solution for $p-q$
Duality in Two Dimensional Gravity. Preprint
UT-582-TOKYO}\refs{\kasch{--}\fkna}
and to describe the moduli space of solutions to this string equation.
The aim of the present paper is to prove similar results for the version of the
string equation arising in the UMM.
It was shown in \nref\abi{Anagnostopoulos, K. N.,
Bowick, M. J.  and Ishibashi, N.:
An Operator Formalism for Unitary Matrix Models.
Mod. Phys. Lett. {\bf A6},  2727-2739 (1991)}\abi\
that the string
equation of the UMM
takes the form $\lb \cp ,\cq_-\rb =\hbox{\rm const.}$, where for the
${\rm k}^{\rm th}$ multicritical point $\cp$ and
$\cq_-$ are $2\times 2$ matrices of
differential operators of order $2k$ and $1$ respectively.
For every solution of the string equation one can construct,
with this result, a pair of points of the $\Gr$ obeying
certain conditions. These conditions lead directly to the Virasoro
constraints for the corresponding $\t$-functions and give a description of the
moduli space of solutions. We stress that the above results depend solely
on the existence of a continuum limit in which
the string equation has the form
$\lb \cp ,\cq_-\rb =\hbox{\rm const.}$ and the
matrices of differential operators $\cp$ and $\cq_-$ have a particular
form to be discussed in detail in subsequent sections. Our results do not
depend on other details of
the underlying matrix model.

The paper is organized as follows. In section 2 we review the double scaling
limit of the UMM in the operator formalism \abi .
Since the square root of the specific heat
flows according to the
mKdV hierarchy we note that its Miura transforms
flow according to KdV and thus give rise to two
$\t$-functions related by the Hirota bilinear equations
of the mKdV hierarchy \nref\peka{
Peterson, D. H.  and Kac, V. G.: Infinite Flag Varieties and Conjugacy
Theorems. Proc. National Acad. Sci. USA {\bf 80}, 1778-1782 (1983)
}\nref\pekac{Kac, V. G.  and
Peterson, D. H.: Lectures on the Infinite Wedge
Representation and the MKP Hierarchy. S\'em. Math. Sup., {\bf vol. 102},
pp. 141-184. Montr\'eal:
Presses Univ. Montr\'eal 1986}\nref\kawa{
Kac, V. G. and Wakimoto, M.: Exceptional Hierarchies of Soliton
Equations. Proc. of Symposia in Pure Math. {\bf 49}, 191-237 (1989)
}\refs{\peka{--}\kawa}.
In section 3 we derive a description of the moduli space of the string equation
in terms of a pair of points in $\Gr$ related by certain conditions. In section
4 we show the correspondence between points in $\Gr$ and solutions to the mKdV
hierarchy. The Virasoro constraints are derived from invariance conditions on
the points of $\Gr$ along the lines of \refs{\kasch , \fknb} .
This is most conveniently done in the
fermionic representation of the $\t$-functions of the mKdV hierarchy. Finally
in section 5 we determine the moduli space of the string equation. It is
found to be isomorphic to the  two fold covering of the
space of $2\times 2$ matrices
$\bigg( P_{ij}(z) \bigg)$, where $P_{ij}(z)$ are polynomials in $z$
such that $\poi$ and $\pio$ are even polynomials having equal degree and
leading terms and
$\poo$ and $\pii$ are odd polynomials of
lower degree satisfying the conditions
$\poo+\pii=0$.

\bigskip
\centerline{\bf 2. The Symmetric Unitary Matrix Model} \bigskip
In this paper we will study the UMM defined by the
one matrix integral
\eqn\partuni{
Z_N^U= \int DU\, {\rm exp}\{ -{N\over \l}\, \Tr \, V(U+U^{\dagger}) \}\, ,
}
where $U$ is a $2N\times 2N$ or a $(2N+1)\times (2N+1) $ unitary matrix,
 $DU$ is the Haar measure for the unitary group and the potential
\eqn\pote{
V(U)=\sum_{k\geq 0}\, g_k\, U^k \, ,
}
is a polynomial in $U$.
As standard
we first reduce the above integral to an
integral over the eigenvalues \nref\brez{Br\'ezin, E.
Itzykson, C., Parisi,  G.  and Zuber, J.B.:
Planar Diagrams.
Commun. Math. Phys. {\bf 59},  35-51 (1978)}\refs{\gros , \brez}
$z_i$ of $U$ which lie
on the unit circle in the complex $z$ plane.
\eqn\partunit{
Z_N^U= \int \lbrace \prod_j
{ {dz_j}\over {2\pi i z_j} } \rbrace
\, | \Delta (z)|^2 {\rm exp}
\{ -{N\over \l}\, \sum_i V(z_i+z^*_i)\} \, ,
}
where $\Delta (z)= \prod\limits_{k < j}\, (z_k-z_j)$
is the Vandermonde determinant. The Vandermonde determinant is conveniently
expressed in terms of trigonometric
orthogonal polynomials \nref\myer{
Myers, R.C.  and Periwal, V.:
Exact Solutions of Critical Self Dual Unitary Matrix Models
Phys. Rev. Lett. {\bf 65},  1088-1091
(1990) }\myer\
\eqn\poly{
 \eqalign{
    c_n^{\pm}(z)
         &=z^n\pm z^{-n}+
        \sum_{i=1}^{i_{max}}\alpha^{\pm}_{n,n-i}(\,z^{n-i}\pm\,z^{-n+i})\cr
         &=\pm c_n^{\pm}(z^{-1})\cr
          }
}
where for $U(2N+1)$ $n$ is a non-negative integer and $i_{max}=n$
and for U(2N) $n$ is
a positive half-integer and $i_{max}=n-{1\o 2}$ . The polynomials
$c_n^{\pm}(z)$ are orthogonal with respect to the inner product
\eqn\inner{
\eqalign {
  \langle c_n^{+},c_m^{+}\rangle
     &= \meas\, \exp \{ -{N\over \l}\, V(z+z^*)\}\, c_n^{+}(z)^* c_m^{+}(z)
\cr
     &={\rm e}^{\phi_n^{+}}\,\delta_{n,m} \, ,\cr
  \langle c_n^{-},c_m^{-}\rangle &={\rm e}^{\phi_n^{-}}\,\delta_{n,m} \, ,\cr
  \langle c_n^{+},c_m^{-}\rangle &=0 \, .\cr
          }
}
The expression for the Vandermonde determinant is
\eqn\vanpoe{
|\Delta(z)|^2=\bigg| det\pmatrix{c_i^-(z_j)\cr c_i^+(z_j)\cr}  \bigg|^2
\quad ,
}
where $j=1,\ldots ,2N$,
$i={1\o 2},{3\o 2},\ldots ,N-{1\o 2}$ for $U(2N)$ and
$j=1,\ldots ,2N+1$,
$i=0,1,\ldots,N$ for $U(2N+1)$
(where the line $c_0^-(z)\equiv 0$ is understood to be omitted).
Then the partition function of the model is given by the product of the norms
of the orthogonal polynomials \bowi\
\eqn\parttau{
Z^U_N=\prod_n {\rm e}^{\phi_n^{+}}{\rm e}^{\phi_n^{-}}=
\tau_N^{(+)}\tau_N^{(-)}\, .
}
In constructing the continuum limit of the UMM
we will also need the orthonormal functions
\eqn\funpi{
\pi_n^{\pm}(z)={\rm e}^{-\phi_n^{\pm} /2 } {\rm e}^{-{N\over {2\l} } V(z_+) }
c_n^{\pm}(z)
}
such that
\eqn\innpi{
\eqalign{
\langle \pi_n^{+}(z),\pi_m^{+}(z) \rangle
                 &=\meas \pi_n^{+}(z)^* \pi_m^{+}(z)\cr
                 &= \delta_{n,m} \, ,\cr
\langle \pi_n^{-}(z),\pi_m^{-}(z) \rangle&= \delta_{n,m} \, ,\cr
\langle \pi_n^{+}(z),\pi_m^{-}(z) \rangle&= 0 \, .\cr
        }
}
The action of the operators $z_{\pm}=z\pm{1\over z}$ and $\zd$
on the $\pi_{n}^{\pm}(z)$ basis is given by finite term recursion relations
\refs{\bowi ,\abi}
\eqn\recplp{
\eqalign{
z_+\, \pi_n^{\pm}(z)&=
\sqrt{R_{n+1}^{\pm} }\pi_{n+1}^{\pm}(z)-r_n^{\pm} \pi_{n}^{\pm}(z) +
\sqrt{R_{n}^{\pm} } \pi_{n-1}^{\pm}(z)~,   \cr
z_-\, \pi_n^{\pm}(z)&=
\sqrt{Q_{n+1}^{\mp} }\pi_{n+1}^{\mp}(z)-q_n^{\pm}
\sqrt{Q_n^{\mp}\over R_n^{\pm}  } \pi_{n}^{\mp}(z) -
\sqrt{Q_{n}^{\pm} } \pi_{n-1}^{\mp}(z)   \, ,   \cr
\zd \pi_{n}^{\pm}(z)&=
-\nl2 \sum_{r=1}^k (v_z^{\pm})_{n,n+r} \pi_{n+r}^{\mp}(z)
+ \biggl \{ n \sqrt{ Q_n^{\mp}\over R_n^{\pm}  }-\nl2 (v_z^{\pm})_{n,n} \biggr
\}
\pi_{n}^{\mp}(z) \cr
       &\, +\nl2 \sum_{r=1}^k (v_z^{\pm})_{n,n-r} \pi_{n-r}^{\mp}(z)~~,\cr
           }
}
where $R_n^{\pm}={\rm e}^{\phi_n^{\pm}-\phi_{n-1}^{\pm} }$,
 $Q_n^{\pm}={\rm e}^{\phi_n^{\pm}-\phi_{n-1}^{\mp} }$,
 $r^{\pm}_n={ {\partial \phi_n^{\pm} }\over {\partial g_1} } $,
 $q_n^{\pm}= { { (Q_{n+1}^{\pm}-Q_n^{\pm})+(R_{n+1}^{\mp}-R_n^{\pm}) }\over
              { r_n^{\pm}-r_n^{\mp} }  }$, and

%
%
%\eqn\matrixoula{
$$(v_z^{\pm})_{n,n-r}=\meas \pi_{n-r}^{\mp}(z) ^*\, \{ \zd V(z_+)\} \,
   \pi_{n}^{\pm}(z) \, ~.
$$
%}
%
The double scaling limit corresponding to the $k^{\rm th}$ multicritical
point is defined by $N\rightarrow\infty$ and $\l\rightarrow\l_c$, with
$t=(1-{n\over N})N^{ {2k}\over {2k+1} }$,
$y=(1-{\l\over\l_c})N^{ {2k}\over {2k+1} }$ held fixed.
It was shown in \abi\ that the operators $z_{\pm}$ and $\zd$
have a smooth continuum limit given by
\eqn\contzform{
\eqalign{
z_+ & \rightarrow 2+ \Ntwo \, {\cal Q}_+ \, ,  \quad
z_-\rightarrow - 2\None \, {\cal Q}_- \, ,  \cr
\zd & \rightarrow N^{ {1\over {2k+1} } } ~{\cal P}_k ~,\cr
        }
}
where ${\cal Q}_{\pm}$ are given by
\eqn\contzp{
\eqalign{
{\cal Q}_-&=
\pmatrix{0&\partial  +v\cr
         \partial  -v&0\cr} \, ,    \cr
{\cal Q}_+&=
\pmatrix{ {(\partial+v)(\pa -v) }&{0}\cr
         {0}&{ (\partial -v)(\pa +v) }\cr } \cr
          &={\cal Q}_-^2\, ,\cr
        }
}
and ${\cal P}_k$ by
\eqn\glob{{\cal P}_k=\left(
\matrix{
        0&{\bf P}_k\cr
{\bf P}^{\dagger}_k&0\cr
       } \right) \, .
}
Here
$\partial\equiv { \partial\over {\partial x} }$ and $x=t+y$.
The scaling function
$v^2$ is proportional to the specific heat $-\pa^2\ln Z$ of the model.
The operators $\bp_k$ are differential operators of order $2k$.
The same assertions hold if we introduce sources $t_{2k+1}$($t_1\equiv x$)
and deform the $k^{\rm th}$ multicritical
potential $V_k$ to $V_k(z)-\sum\limits_l t_{2l+1} V_l(z) N^{2(k-l)\o 2k+1}$.
{}From $\lb \zd , z_- \rb=z_+$ it follows that
\eqn\strcon{
\lb {\cp}_k ,\cq _- \rb =1 \, ,
}
where $\cq_-$ has the form \contzp\ and ${\cp}_k$ has the form \glob.
We stress here that this equation holds for the system perturbed
away from the multicritical points as well as exactly at
multicriticality.
Our main aim is to study equation \strcon\ -
the string equation for the UMM.

For completeness we will present here some information about the
solutions of \strcon\ that was obtained in \abi\ (or follows from
the same analysis). Most of these facts will also follow
from the results of Sections 3-5; the reader may go directly to these
sections.

It is proved in \abi\ that ${\bf P}_k$ are given at the $k^{\rm th}$
multicritical point by
\eqn\pko{
\bp_k=\bpt_k - x\, ,
}
where
\eqn\pk{
\bpt_k=a_k^{-1}\{ (\pa+v)\lb (\pa-v)(\pa+v) \rb ^{k-1/2}\} _+ \, ,
}
and $a_k^{-1}=2(2k+1)\sum\limits_{l=1}^k (-1)^l ~l^{2k} { B(k+1,k+1) \over
\Gamma(k-l+1)\Gamma(k+l+1)}$.
Here $\Psi_+$ denotes the differential part of a pseudodifferential
operator $\Psi$.  One can give the corresponding expression
$\bp=-\sum\limits_{l\geq 1}(2l+1)t_{2l+1}\bpt_l-x$
for perturbations from the $k^{\rm th}$
multicritical point. These expressions
can be used to get an ordinary differential
equation for the specific heat $v$ in the form
\eqn\streq{
{\hat{\cal D}} \R _{k}\lb u \rb =a_k vx \, ,
}
where ${\hat{\cal D}}=\partial + 2v$, $u=v^2-v'$, and
$\R _{k}\lb u \rb $ are the
Gel'fand-Dikii potentials defined through the recursion relation
\eqn\kdvpot{
\pa\, \R _{k+1}\lb u \rb =\biggl( {1\o 4}\pa ^3-{1\o 2}(\pa u+u\pa) \biggr)
\, \R _{k}\lb u \rb \, ,\quad
\R _{0}\lb u \rb ={1\over 2} \, .
}
In the non-critical model the analogous equation is
\eqn\perstr{
\sum_{l\geq 1}(2l+1)t_{2l+1}{\hat{\cal D}} \R _{l}\lb u \rb =-vx \, .
}
The equation $\lb \zd , z_+ \rb=z_-$ in the continuum limit becomes
$\lb {\cp}_k ,\cq _+ \rb =2\cq_-$ and is consistent with the relation
$\cq_-^2=\cq_+$.

Equation \streq\ is closely related to the mKdV hierarchy. Indeed, by
slightly modifying the calculations of \refs{\cdm ,\holo}, one
can show that $v$ flows
according to the mKdV hierarchy
\eqn\mkdvfl{
{ {\partial v}\over {\partial t_{2k+1} } }=-\partial
{\hat{\cal D}}  \R _{k}\lb u \rb\, .
}
By introducing scaling
operators
\eqn\scop{
\langle \sigma_k \rangle= {\partial \over {\partial t_{2k+1}}} ln \, Z
}
one can show that
\eqn\scpot{
\langle \sigma_k\sigma_0\sigma_0 \rangle=
2v \pa{\hat{\cal D}} \R _{k}\lb u \rb \, .
}
Then $\langle \sigma_0\sigma_0 \rangle= -v^2 $ and
$ \langle \sigma_k\sigma_0\sigma_0 \rangle=
{\partial\over{\partial t_{2k+1} } }
\langle \sigma_0\sigma_0 \rangle$ imply equation
\mkdvfl.

If $v$ flows according to mKdV, then  the functions $u_1=v^2+v'$ and
$u_2\equiv u=v^2-v'$ will flow according to KdV, being related to $v$ by the
Miura transformation. The flows of $u_1$ and $u_2$ have associated
$\t$-functions $\to$ and $\tt$ such that
\eqn\deftau{
u_1=-2\partial^2\, ln\, \to\, , \qquad
u_2=-2\partial^2\, ln\, \tt\, .
}
Then
\eqn\protau{
v^2=-\pa ^2\, ln\,(\to\tt)\, ,\quad
v=\partial\, ln\, {\tt \over \to}\,
}
The Miura transformation $u_1=v^2+v'$ yields the
simplest bilinear Hirota equation
of the mKdV hierarchy \refs{\peka{--}\kawa}, namely
\eqn\reltau{
D^2\, \to\cdot\tt =\to ''\tt-2\to '\tt '+\to \tt ''=0
}
where $D$ denotes the Hirota derivative.
The structure of this hierarchy will be examined further in section 4.
Note that \protau\ shows that the partition
function $Z$ of the UMM is given
by
\eqn\miwa{
Z=\to\cdot\tt
}
with the two mKdV $\t$ functions being related by
\reltau\ .
\bigskip
\bigskip
\centerline{\bf 3. The Sato Grassmannian}
\bigskip
The partition function of the
UMM was shown in Section 2 to be the product
of two mKdV $\t$-functions $\to $ and $\tt$. As will be explained in Section 4,
any $\t$ function that can be represented by
a formal power series corresponds
to a point of the big cell of the Sato Grassmannian $Gr^{(0)}$.
It will  be shown that the mKdV flows can be described by the
flows of two points $V_1 ,\, V_2\in\Gr$ that are related by certain
conditions preserved by the flows. The string equation will impose
further conditions that will pick out a unique pair $(V_1 ,V_2)$. It will
further impose constraints on the $\t$-functions, which turn out to be the
expected Virasoro constraints \refs{\cdm , \holo}.
The treatment described here follows
closely that for the case of the HMM
\refs{\aschw{--}\fkna}.

Consider the space of formal Laurent series
$$H=\{ \sum\limits_n a_n z^n\, ,\quad a_n=0\quad\hbox{for}\quad n\gg 0\,\}$$
and its decomposition
$$H=H_+ \oplus H_- \, ,$$
where
$H_+=\{ \sum
\limits_{n\geq 0} a_n z^n\, ,\quad a_n=0\quad\hbox{for}\quad n\gg 0\,\}$.
Then the big cell of the Sato Grassmannian $\Gr$ consists of all subspaces
$V\subset H$ comparable to $H_+$, in the sense that the natural projection
$\pi_+ :\, V\rightarrow H_+$ is an isomorphism.

Consider the space $\Psi$ of pseudodifferential operators
$W=\sum
\limits_{i\leq k}w_i(x)\pa^i$ where the functions $w_i(x)$ are taken to be
formal power series
(i.e. $w_i(x)=
\sum\limits_{k\geq 0}w_{ik}x^{k}\, ,\,\, w_{ik}=0\, ,\,k\gg 0$). $W$ is
then a pseudodifferential operator of order $k$. It is called monic if
$w_k(x)=1$ and normalized if $w_{k-1}(x)=0$. The space $\Psi$ forms an algebra.
The space of monic, zeroth-order pseudodifferential operators forms a group
$\cal G$.

There is a natural action of $\Psi$ on $H$ defined by
$$
\eqalign{
x^m\pa^n : H&\rightarrow H \cr
        \phi&\rightarrow (-{d\o dz})^m(z)^n\, \phi\, .\cr
        }
$$
Then it is well known \nref\mula{The most appropriate exposition for our
purposes is given in Mulase, M.:
Category of Vector Bundles On Algebraic Curves and Infinite Dimensional
Grassmannians. Int. J. Math.
{\bf 1},  293-342 (1990)}\mula\ that
every point $V\in\Gr$ can be uniquely represented in
the form $V=SH_+$ with $S\in{\cal G}$. This will imply that for every operator
$\cq_-$ we can uniquely associate a pair of points $V_1 ,\, V_2\in\Gr$.

Indeed, consider $\so$ and $\st \in{\cal G }$ such that
\eqn\diagq{
{\hat S}\cq_-{\hat S^{-1}}=\cqt_-
}
where
\eqn\diagm{
{\hat S}=\pmatrix{\so&0\cr 0&\st\cr}\, ,\,
\cqt_-=\pmatrix{0&\pa\cr \pa&0\cr}\, .
}
Then
\eqn\condd{
\eqalign{
\so(\pa+v)\sti&=\pa\, ,\cr
\st(\pa-v)\soi&=\pa\, ,\cr
        }
}
which imply that
\eqn\condk{
\eqalign{
\so(\pa^2-u_1)\soi&=\pa^2\quad u_1=v^2+v'\, ,\cr
\st(\pa^2-u_2)\sti&=\pa^2  \quad   u_2=v^2-v'\, .\cr
        }
}
The existence of $\so\in{\cal G}$ follows from the general
fact \nref\sewi{Segal, G.  and Wilson, G.:
Loop Groups and Equations of KdV Type.
Publ. I.H.E.S. {\bf 61}, 5-65 (1985) }\sewi\
that for every monic normalized pseudodifferential operator ${\cal L}$
of order $n$ there exists an $S$ such that $S{\cal L}S^{-1}=\del^n$.

Given $\so$, one can determine $\st$ from
$$\so(\pa+v)=\pa\st.$$

By taking formal adjoints of \condd\ and \condk, it is easy to show that
$\so$ and $\st$ be
made simultaneously unitary. Indeed, from \condk\ we obtain
\eqn\suni{
\eqalign{
(\soi)^{\dagger}(\pa^2-\ut)\sod &=\pa^2\Rightarrow\cr
(\so\sod)^{-1}\pa^2(\so\sod) &=\pa^2\Rightarrow\cr
\so\sod&=f(\pa^2)\, ,\cr
        }
}
where $f$ is arbitrary. Similarly $\st\std=g(\pa^2)$. But since \diagq\ implies
\eqn\prfun{
({\hat S}{\hat S^{\dagger}})^{-1}
\pmatrix{0&\pa\cr\pa&0\cr}
({\hat S}{\hat S^{\dagger}})=
\pmatrix{0&\pa\cr\pa&0\cr}\, ,
}
then
$$
({\hat S}{\hat S^{\dagger}})=
\pmatrix{f(\pa^2)&0\cr 0&g(\pa^2)\cr}
$$
gives
$$
\eqalign{
\pa g&=f\pa\, ,\cr
\pa f&=g\pa\, ,\cr
        }
$$
or, $f=g$. Therefore $\so$ and $\st$
can be simultaneously chosen to be unitary,
i.e $\so\sod=1$ and $\st\std=1$.

Since $V\subset\Gr$ is given uniquely by $V=SH_+$, the operator $\cq_-$
determines two spaces $\vo=\so H_+$ and  $\vt=\st H_+$. Conversely given
spaces $\vo$ and $\vt$ determine
$\cq_-$ uniquely. The operator $\cq_-$, however,
is a differential operator  and $\vo,\vt$
cannot be arbitrary. Indeed, since every differential operator leaves $H_+$
invariant, we obtain
\eqn\incond{
\eqalign{
(\pa+v)\,H_+\subset H_+
\Leftrightarrow&\soi\pa\st\, H_+\subset H_+\cr
\Leftrightarrow&\pa\, \vt\subset\vo\cr
\Leftrightarrow&z\, \vt\subset\vo\cr
        }
}
Similarly, $z\, \vt\subset\vo$.

The string equation will impose further conditions on $\vo$ and $\vt$.
After transformation with the operator $\hat{S}$
equation \strcon\ becomes
\eqn\strdia{
\lb \cpt_{(k)} ,\cqt_- \rb =1
}
where $\cpt_{(k)}={\hat S}\cp_{(k)}{\hat S^{-1}}$. The solution to \strdia\ is
\eqn\psoln{
\cpt_{(k)}=\pmatrix{ 0& -x+{\tilde f}_k(\pa)\cr -x+{\tilde f}_k(\pa)&0\cr}
}
which gives $\bp_{(k)} =\soi \big(-x+{\tilde f}_k(\pa)\big)\st$ and
$\bpd_{(k)} =\sti \big(-x+{\tilde f}_k(\pa)\big)\so$.
Consistency requires therefore
that $-x+{\tilde f}_k(\pa)$
must be self adjoint ${\tilde f}_k(\pa)=f_k(\pa^2)$.
For the $k^{\rm th}$ multicritical
point $\bp_{(k)}$ is a differential operator of order $2k$. Therefore
$f_k(\pa^2)=\pa^{2k}+\ldots$. By using the freedom to redefine $S_i$ by a
monic,
zeroth-order,  pseudodifferential
operator $R=1+\sum\limits_{i\geq 1}r_i\pa^{-i}$ with constant
coefficients $r_i$, it is
easy to show that all negative powers in $f_k(\pa^2)$ may be eliminated. The
proof shows that all powers below $\pa^{-1}$ can be eliminated by $R$,
and a $\pa^{-1}$ term is forbidden by self-adjointness. Therefore
\eqn\formf{
f_k(\pa^2)=\pa^{2k}+\sum\limits_{1\leq i\leq k}f_i(x)\pa^{2(k-i)}
}
By Fourier transforming, the action of $\cpt$ on $H$ is represented by
\eqn\fourp{
\cpt_{(k)}=\pmatrix{0&A_k\cr A_k&0\cr}\, ,\hbox{\rm where}\quad
A_k={d\o dz}+\sum\limits_{i=0}^k \a_i z^{2i}
\quad\hbox{\rm and $\a_i=$const.}
}

Given the constants $\a_i$, we can calculate
the operator $\bp_{(k)}$. Since
$\st(\pa -v)(\pa+v)\sti=\pa^2$ implies
$\st\lb(\pa-v)(\pa+v)\rb^{i-{1\o 2}}\sti=\pa^{2i-1}$ then using
$\so(\pa+v)\sti=\pa$ we obtain
\eqn\intca{
\so(\pa+v)\lb(\pa-v)(\pa+v)\rb^{i-{1\o 2}}\sti=\pa^{2i}\, .
}
Transforming back to $H_+$ we obtain
\eqn\mapa{
\eqalign{
\bp_{(k)}&=\soi(-x+\sum_{i=0}^k\a_i\pa^{2i})\st \cr
         &=\soi(- x+\a_0)\st+\sum_{i=1}^k\a_i\soi\pa^{2i}\st \cr
         &=\soi(- x+\a_0)\st+\sum_{i=1}^k\a_i
            (\pa +v)\lb(\pa-v)(\pa+v)\rb^{i-{1\o 2}}\cr
        }
}
Comparing with \pk\ and since
$\soi x\st=x+\sum\limits_{i\ge 1}q_i(x)\pa^{-i}$,
we conclude that at the ${\rm k}^{\rm th}$ multicritical point, $\a_k=1$ and
$\a_i=0$ for $i<k$. Moreover, by perturbing away from the multicritical points
we see that
\eqn\mevol{
\a_i(t)=-(2i+1)t_{2i+1}\, .
}

The requirement that $\cp$ be a differential operator
is equivalent to the conditions
$A_k\,\vo\subset\vt$ and $A_k\,\vt\subset\vo$.
The space of solutions
to the string equation is the space of operators $\cq_-$ such that
there exists $\cp_{(k)}$ with $\lb \cp_{(k)} ,\cq_- \rb=1$.
We conclude that this space is isomorphic to the
set of elements $\vo ,\vt\subset\Gr$ that satisfy the conditions:
\eqn\fincon{
\eqalign{
z\,\vo\subset\vt \quad z\,\vt\subset\vo\cr
A_k\,\vo\subset\vt \quad A_k\,\vt\subset\vo\cr
        }
}
for some $A_k={d\o dz}+\sum\limits\limits_{i=0}^k \a_i z^{2i}$.

It is now easy to show that the string equation is
compatible with the mKdV flows \mkdvfl . We will show
in the next section that the mKdV flows for the
scaling function $v$ are equivalent to the condition
\eqn\spaflo{
{\pa \o\pa t_{2k+1}}\, V_i =z^{2k+1}\, V_i\quad (i=1,2)\, .
}
Then $V_i(t)=\exp \{ \sum\limits_k t_{2k+1}z^{2k+1} \}V_i\equiv \gamma(t,z)V_i$
and \fincon\ imply
\eqn\cons{
\eqalign{
z\,\gamma(z,t)V_1\subset\gamma(t,z)V_2&
       \Rightarrow z\,V_1(t)\subset V_2(t)\cr
A_k(t)\,\gamma(z,t)V_1\subset\gamma(t,z)V_2&
       \Rightarrow A_k(t)\,V_1(t)\subset V_2(t)\, ,\cr
        }
}
where
\eqn\newop{
A_k(t)\equiv\gamma A_k \gamma^{-1}=A_k-\sum\limits_k(2k+1)t_{2k+1}z^{2k}
}
and analogous equations with $V_1$ and $V_2$ interchanged.
This is clearly consistent with \mevol .

{}From \fincon\ we see that $z^2$, $zA$ and $A^2$ leave $V_{1,2}$ invariant.
In the next section we show that this fact implies
Virasoro constraints for the $\t$-functions associated with the mKdV
flows of the UMM.
\bigskip
\centerline{\bf 4. The mKdV $\t$-functions and the Virasoro constraints}
\bigskip
In this section we will describe the $\t$-function formalism for the mKdV
system and give a derivation of the Virasoro constraints on the $\t$-functions
of the UMM. These will be derived from the invariance conditions \fincon\ on
the spaces $\vo$ and $\vt$ following the lines of
\refs{\kasch ,\fknb } for the
HMM. The idea is to transform the Virasoro generators into
fermionic operators in the fermionic representation of $\GL$ using the
boson-fermion equivalence. Then using the correspondence between
$\GL$-orbits of the vacuum and $\Gr$, annihilation of the
$\t$-function by the Virasoro constraints $\L_n$
is shown to be equivalent to the
invariance of $V\in\Gr$ under the action of operators $z^{2n}A_{KdV}$.
In \refs{\aschw, \schwa}, it was
shown that $A_{KdV}$ was nothing but the operator ${\rm P}$
of the HMM acting on $\Gr$, and the Virasoro constraints were proved from the
string equation. We summarize below these results and derive the Virasoro
constraints for the UMM from the conditions \fincon .

First we introduce the fermionic representation of $\GL$ on the Fock space F of
free fermions. The fermionic operators are defined to satisfy the
anticommutation relations
\eqn\fercom{
\{\ps_i,\psd_j\}=\d_{ij}\, ,\quad\{\ps_i,\ps_j\}=\{\psd_i,\psd_j\}=0
\quad (i\in\Z)\, .
}
The vacuum $|0>$ satisfies
\eqn\vacan{
\ps_i|0>=0\,\,\quad{\rm for}\,\,\quad i>0\, ,\quad\quad
\psd_i|0>=0\,\,\quad{\rm  for}\,\,\quad i\leq 0\, ,
}
and the states ( $m>0$)
\eqn\chgst{
|m>=\psd_m\ldots\psd_1|0>\, ,\quad |-m>=\ps_{-m+1}\ldots\ps_0|0>\,
}
are the filled states with charge $m$ and $-m$ respectively. The operators
$\psd_i$ and $\ps_i$ have been assigned charges $1$ and $-1$ respectively and
the vacuum $|0>$ charge $0$. The normal ordering is defined by
\eqn\orde{
:\psd_i\ps_j:\,\,=\psd_i\ps_j-<\psd_i\ps_j>=
\cases{\psd_i\ps_j\quad i>0\cr
      -\ps_j\psd_i\quad i\leq 0\cr}
}
Then the fermionic representation of the algebra $\gl$ is defined by
\footnote{$^1$}{ Note that this representation of $\gl$ and
$\GL$ is equivalent to the infinite wedge representation \pekac .}
\eqn\frepa{
r_F(a)|\chi>=\sum_{i,j}:\psd_i a_{ij} \ps_j:|\chi>\,\quad a\in\gl\quad
|\chi>\in \F
}
and of the group $\GL$ by
\eqn\frepg{
\eqalign{
R_F(g)&\Big(  \psd_{i_1}\psd_{i_2}\ldots\ps_{i_1}\ps_{i_2}\ldots\Big) |-m>=\cr
      &\Big( (\psd g)_{i_1}(\psd g)_{i_2}\ldots
        (g\ps)_{i_1}(g\ps)_{i_2}\ldots\Big)|-m>\cr
        }
}
for $m\gg 0$ such that $(\psd g)_{-j}=\psd_{-j}$ for $j>m$. In \frepg ,
$g\in\GL$ and
$(\psd g)_{i}\equiv\psd_jg_{ji}$ and $(g\ps)_i\equiv g_{ij}\ps_j$. The above
representation conserves the charge and therefore preserves the decomposition
$$
\F=\oplus_{m\in\Z}\F^{(m)}
$$
where $\F^{(m)}$ is the space of states with charge $m$. The first step in
order to establish the boson-fermion correspondence is to define the current
operators
\eqn\curdef{
J_n=\sum_{r\in\Z}:\psd_{n-r}\ps_r:\quad n\in\Z
}
which satisfy the bosonic commutation relations
\eqn\comcur{
\lb J_m,J_n\rb=m\d_{m,-n}\, .
}
Then we define an isomorphism $\sigma :\F\rightarrow\B$ where
the bosonic Fock space
$\B=\oplus_{m\in\Z}
\B^{(m)}\cong\C\lb t_1,t_2,\ldots,;u,u^{-1}\rb$ of polynomials in
$t_1,t_2,\ldots,;u,u^{-1}$ by the requirement
\eqn\isoop{
\sigma\big( |m> \big)=u^m\, ,\quad
\sigma J_n\sigma^{-1}={\pa\o \pa t_n}\,\, (n\ge 0) \quad
\sigma J_n\sigma^{-1}=-nt_{-n}\,\, (n< 0) \, .
}
Then the state $|\chi>\in\F$ is represented in $\B$ by
\eqn\fndeft{
\t^{\chi}(t;u,u^{-1})=
\sum_{m\in\Z}u^m<m|{\rm e}^{\sum_{p\ge 1}t_pJ_p}|\chi>\equiv
\sum_{m\in\Z}u^m\t^{\chi}_m(t)
}
Note that $\sigma=\oplus_{m\in\Z}\sigma_m$, where
$\sigma_m :\F^{(m)}\rightarrow\B^{(m)}\cong u^m\C\lb t_1,t_2,\ldots\rb$
and
$\t(t)=\oplus_{m\in\Z}\t_m(t)$.

Then one observes that if the state $|g>_0$ belongs to the $\GL$ orbit of the
vacuum (i.e. $|g>_0\quad=g|0>$ for some $g\in\GL$), then
$\sum\limits_{j\in\Z}\psd_j|g>_0\otimes\ps_j|g>_0\quad = 0$
leads to the bilinear Hirota equations for the $\t$-functions of the KP
hierarchy (see \refs{\peka{--}\kawa} for details). The KP
$\t$-function belongs to the $\GL$ orbit of
the vacuum and is given by
\eqn\tauorb{
\t=<0|{\rm e}^{\sum_{p\ge 1}t_pJ_p}g|0>\,\,\in\GL\cdot 1\, .
}

Similar considerations apply for the $k^{\rm th}$ modified KP (mKP) hierarchy.
This is defined by the equation
$\sum\limits_{j\in\Z}\psd_j|g>_k\otimes\,\ps_j|g>_0\, =0$
where $|g>_k$ belongs to the $\GL$ orbit of the state $|k>$ of \chgst .
Kac and Peterson \peka\ showed that this is equivalent to the mKP $\t$-function
$\t(t)=\t_k(t)\oplus\t_0(t)$ lying on the $\GL$ orbit of
$|k>\oplus\, |0>$.

One can go further and observe that the Kac-Moody algebra of $sl_n$
(thought of as $\hat{sl_n}\big(n,\C\lb u,u^{-1}\rb \big)$) when embedded
in $\gl$
%
%\footnote{$^2$}{ More precisely in
%$\bar{a}=\{ (a_{ij})\, i,j\in\Z|
%\hbox{\rm for each k the number of non-zero $a_{ij}$ with $i\leq k$
%and $j\ge k$ is finite }
%         \}$
%}
%
has irreducible highest weight representations on the space
$\B_{(n)}=\oplus^{n-1}_{m=1}\B^{(m)}_{(n)}$
where
$\B^{(m)}_{(n)}=\C\lb t_j|j\not= 0\, {\rm mod}\, n\rb\subset\B^{(m)}$.
Therefore one can restrict the mKP(resp. KP) hierarchies and obtain the so
called n-reduced mKP(resp. KP) hierarchies. Then one can show \peka\ that the
$\t$-function
$\t_{(n)}=\oplus_{k=0}^{n-1}\t_k$
belongs to the $\hat{SL_n}$ orbit of the sum of the highest weight vectors
$\oplus_{m=0}^{n-1}1_m$. We are mainly interested in the second reduced
mKP hierarchies. Then the simplest bilinear Hirota equations give for
$u_i=-2\pa^2\ln\t_i\, ,i=1,2$ and $v=\ln{\tt\o\to}$ equations
\deftau\ and \protau , and we obtain the mKdV hierarchy.

Now we want to establish the relation between elements of $\Gr$ and fermionic
states. Consider $V\in\Gr$ spanned by the
vectors $\{\phi_i\}\, (i=0,1,2,\ldots)$
where $\phi_i=\sum\limits_{k\in\Z}\phi_{i,k}z^{k}\in H$. Associate to every
$\phi_i\in V$ a fermionic operator $\psd[\phi_i]$ by
\eqn\fermba{
\psd[\phi_i]=\sum_{k\in\Z}\phi_{i,k}\psd_k\,
}
and to every $V\in\Gr$ the state $|v>$ belonging to the
$GL(\infty)$ orbit of the vacuum and such that
\eqn\vecmap{
\psd[\ph_i]|v>=0\quad\forall i\, ,
}
where $V$ is spanned by the functions $\{\phi_i\}$.
Then because bilinear fermionic
operators
\eqn\biop{
\hat{a}=\sum_{i,j}:\psd_ia_{ij}\ps_j :
}
satisfy
\eqn\braop{
\lb \ps_i,\hat{a}\rb=\sum_k a_{ik}\ps_k\, ,\quad
\lb \hat{a},\psd_i\rb=\sum_k \psd_k a_{ki}\, ,
}
we can associate to them operators $a$ acting on $H$ by
\eqn\boop{
a\, h(z)=\sum_k\bigg(\sum_i a_{ki}h_i\bigg) z^{k}\,\quad
(h(z)\in H)\, .
}
Then if
\eqn\eqco{
\eqalign{
\hat{a}_1\leftrightarrow a_1\quad &\hbox{\rm and}
                           \quad\hat{a}_2\leftrightarrow a_2
\quad\hbox{\rm then}\cr
\lb \hat{a}_1, \hat{a}_2\rb &\leftrightarrow
                    \lb a_1,a_2\rb\, .\cr
        }
}
Moreover, one can prove \refs{\kasch,\fknb} that if $|v>$
corresponds to $V\in\Gr$, then
\eqn\relop{
\hat{a}|v>={\rm const.}|v>\Leftrightarrow a\, V\subset V\, .
}
The proof follows immediately from the remark that
$[\hat{a},\psi^{\dagger}(\phi)]=\psi^{\dagger}(a\phi)$ (see \braop).
Thus if $\hat{a}|v>={\rm const.}|v>$ and $\phi\in V$ i.e.
$\psi^{\dagger}(\phi)|v>=0$, then
$\psi^{\dagger}(a\phi)|v>=(\hat{a}\psi^{\dagger}
(\phi) - \psi^{\dagger}(\phi)\hat{a})|v>=0$ and hence $a\phi\in V$.
In other words $a\, V\subset V.$ In a similar way
one can establish the implication in \relop\ in the reverse
direction.
{}From the above discussion we see that if $V_{1,2}$ are to describe mKdV flows
then they should correspond to states
$|v_1>\in \GL\cdot|0>$ and $|v_2>\in \GL\cdot|1>$. Then since
$|v_i>_t=\exp \{ \sum\limits_{p\ge 1}t_{p}J_{p} \}|v_i>$
or
\eqn\feflo{
{\pa\o \pa t_{2k+1}}|v_i>_t=J_{2k+1}|v_i>_t\, ,
}
equation \boop\ yields \spaflo .

Consider the Virasoro operators
\eqn\virns{
\L_n={1\o 2}\sum_{p=-\infty}^{2n-1}J_pJ_{2n-p}+{1\o 16}\d_{n,0}
\quad n\ge 0
}
acting on the $\t$-functions associated with the states $|g>_i$
\eqn\nstf{
\t_i(t)=<i-1|\exp \{\sum\limits_{p\ge 1}t_pJ_p\}|g>_i\quad i=1,2\, .
}
Then shift the times
%$\{ t_1,t_2,\ldots \}\rightarrow
%\{ t_1+\a_0,t_2,t_3+{\a_1\o 3},\ldots,t_{2k+1}+{\a_k\o 2k+1},t_{2k+2},
%t_{2k+3},\ldots \} $
$t_{2i+1}\rightarrow t_{2i+1}+{\a_i\o 2i+1}$ for $i\leq k$,
where the $\a_i$ are defined in \fourp . Then
\eqn\shftl{
\eqalign{
\t_i(t)\rightarrow\t_i'(t)&=
             <i-1|\exp \{\sum\limits_{p\ge 1}(t_p+t_p^{(0)})J_p \}|g>_i\, ,\cr
\L_n\rightarrow \L_n'&=
{\rm e}^{\sum\limits_{p=0}^{k}{\a_p\o 2p+1}J_{2p+1}}
          \L_n{\rm e}^{-\sum\limits_{p=0}^{k}{\a_p\o 2p+1}J_{2p+1}}\cr
                     &=
\L_n+\sum_{p=0}^{k}\a_p J_{2(n+p)+1}\, .\cr
        }
}
In \refs{\kasch ,\fknb } it
was shown that the fermion operators $\L_n'$ correspond via \boop\ to
the operators
\eqn\virgr{
{1\o 2}z^{2n+1}A={1\o 2}z^{2n+1}\left( {d\o dz}+\sum_{p=0}^{k}\a_iz^{2i}
%-{1\o 2}z^{-1}
\right)\, .
}
Then, because of \relop , invariance of $V_{1,2}$ under $z^{2n+1}A$ (see
\fincon\ ) implies that the $\t$-functions $\t_i$ are annihilated by the
$\L_n$'s for $n\ge 1$ and
\eqn\zerom{
\L_0\t_i=\mu\t_i\, .
}
The constant $\mu$ is an arbitrary parameter. Such a parameter
does not appear for $L_n (n\ge 1)$ by closure of the Virasoro algebra.
As pointed out in
\holo\ it is the same for the two $\t$-functions and it
cannot be determined by the
closure of the algebra since, contrary to the HMM, $\L_{-1}$ is absent. If one
includes boundary conditions then there exists a one parameter family of
solutions to the string equation with the correct scaling behaviour at infinity
\nref\watt{
Watterstam, A.:
A Solution to the String Equation of Unitary Matrix Models.
Phys. Lett. {\bf B263},  51-58 (1991)}\watt .
It has been suggested in \holo\ that the parameter of such a particular
solution is related to $\mu$.
The Virasoro constraints are then those of a heighest weight state of conformal
dimension $\mu$.
Although $\L_{-1}$ is absent one should bear in mind the
additional constraints arising from the interrelation of $\to$ and $\tt$
determined by equation \fincon .
\bigskip
\centerline{\bf 5. Algebraic Description of the Moduli Space}
\bigskip
In this section we attempt to give a complete description of the moduli space
of the string equation \strcon . As already mentioned, the space of
solutions to \strcon\
is isomorphic to the set of points $\vo,\vt$ of $\Gr$ that satisfy the
conditions \fincon . Therefore we will start by describing the spaces
$\vo,\vt$.

First choose vectors $\ph_1(z),\ph_2(z) \in\vo$, such that
$$
\ph_1(z)=1+\hbox{lower order terms}\, ,\quad
\phi_2(z)=z+\hbox{lower order terms}
$$
Then the condition $z^2\,\vo\subset\vo$ and $\pi_+(\vo)\cong H_+$
shows that we can choose a basis for $\vo$
$$
\ph_1,\ph_2,z^2\ph_1,z^2\ph_2,\ldots
$$
Since $z\,\vo\subset\vt$
and $\pi_+(\vt)\cong H_+$ we can choose a basis for $\vt$ to be
$$
\ps,z\ph_1,z\ph_2,z^3\ph_1,z^3\ph_2,\ldots
$$
where $\ps (z)=1+\hbox{lower order terms}$. Using $z\,\vt\subset\vo$ we have
$z\ps=\alpha \ph_1+\beta\ph_2$. Choose $\ph_1 ,\ph_2$ such that
$z\ps=\ph_2$. Then we obtain the following basis for $\vo,\vt$
($\ph\equiv\ph_1$):
\eqn\basis{
\eqalign{
\vo \,:\quad &\ph,z\ps,z^2\ph,z^3\ps,\ldots\cr
\vt \,:\quad &\ps,z\ph,z^2\ps,z^3\ph,\ldots\cr
        }
}
Then it is clear that $\ph,\ps$ specify the spaces $\vo,\vt$. Using the
conditions $A\vo\subset\vt$ and $A\vt\subset\vo$ we obtain
\eqn\stdif{
\eqalign{
({d\o dz}+f_k(z^2))\ph&=\poo\ph+\poi\ps\cr
({d\o dz}+f_k(z^2))\ps&=\pio\ph+\pii\ps \, .\cr
        }
}
The polynomials $\poo$ and $\pii$
are odd whereas $\poi,\pio$ are even. Comparing
both sides of \stdif\ we find that because
${\rm deg(f_k)=2k}$, ${\rm deg}(\poi)=deg(\pio)=2k$ and
${\rm deg}(\pii),\, {\rm deg}(\poo)<2k$
and that the coefficients of the leading terms of $\poi$ and $\pio$ are
equal to $\a_k$.

Equations \stdif\ can be rewritten in the form
\eqn\basdif{
D\chi=B_{2k}(z)\chi
}
where $\chi={\ph\choose\ps}$,
\eqn\mabdi{
D=\pmatrix{ {d\o dz}&0\cr 0&{d\o dz} }\, ,\quad
B_{2k}(z)=\pmatrix{ {\poo-f_k(z^2)}&\poi\cr
                    \pio&{\pii-f_k(z^2)}\cr } \, .
}
The requirement that $\ph,\ps$ be solutions of the form
$1+(\hbox{lower order terms})$,
rather than exponential, puts further constraints on the matrix
$B_{2k}(z)$. It requires that the eigenvalues $\l (z)$ of $B$ must vanish up to
${\cal O}(z^{-2})$, i.e
$\l(z)=\sum\limits_{i\ge 1}\l_iz^{-i-1}$. Indeed then
$\chi\sim\exp{\int^{z}\l (z')dz'}\sim\exp{-{\l_1\o z} }\sim
1-\l_1z^{-1}+\ldots$, as desired. But then
${\rm detB}_{2k}(z)$ is of ${\cal O}(z^{-4})$ and
\eqn\detf{
f_{2k}(z^2)={1\o 2}(\poo+\pii)\pm
\sqrt{ {1\o 4}(\poo+\pii)^2-\Delta+{\cal O}(z^{-4})  }
}
where $\Delta(z)=\poo\pii-\poi\pio $.
Since $f(z^2)$ is an even function of $z$,
the odd parity of $\poo$ and $\pii$ determine that $\poo+\pii=0$.

Conversely, given a $2\times 2$ matrix $\bigg( P_{ij}(z)\bigg)$ with
$\poi ,\pio$ even polynomials of degree $2k$ and
$\poo ,\pii$ odd polynomials of degree $<2k$ such that
$\poo+\pii=0$, we will show that we obtain exactly two solutions to the
string equation \strdia . The eigenvalues $\l^{(1,2)}(z)$
of $\bigg( P_{ij}(z)\bigg)$ are given by
\eqn\evl{
\l^{(1,2)}(z) =\pm\sqrt{-\Delta (z)}
}
and $\l^{(i)}(z)=\sum\limits_{j=-\infty}^k\l^{(i)}_jz^{2j}\quad (i=0,1)$. Then
the matrix $B_{2k}$ of \mabdi\ with
\eqn\asyf{
f_k^{(i)}(z^2)=\sum^k_{m=-\infty}\a_m^{(i)}z^{2m}\quad
\a_m^{(i)}-\l_m^{(i)}=
\cases{0&$m\ge 0$\cr\not= 0& at least for $0\gg m$\cr}\,
}
will have determinant at most of ${\cal O}(z^{-4})$. Then the system \stdif\
will have solutions $\phi(z)$ and $\psi(z)$ of the form
$\phi (z),\,\psi(z)={\rm const.}+\hbox{\rm lower order terms}$.
We can set the constant to one by requiring that the leading terms of the
polynomials $\poi$ and $\pio$ are equal.
Since we know
from the discussion at the end of section 3 that
the $m<0$ terms of the operator
$A$ can be gauged away, we see that each eigenvalue $\l^{(i)}(z)$
specifies a unique solution to the string equation \strdia .

Hence the space of solutions to the string equation \strcon\ is the two fold
covering of the space of matrices $\bigg( P_{ij}(z)\bigg)$ with polynomial
entries in $z$
such that $\poi$ and $\pio$ are even polynomials having equal degree and
leading terms and
$\poo$ and $\pii$ are odd polynomials satisfying the conditions
$\poo+\pii=0$ and ${\rm deg}\poo<{\rm deg}\poi$.

\bigskip\bigskip
\bigskip
\centerline{\bf Acknowledgements}
\bigskip

The research of K.A. and M.B.
was supported by the Outstanding Junior Investigator Grant DOE
DE-FG02-85ER40231, NSF grant PHY 89-04035 and a Syracuse University Fellowship.
A.S. would like to thank Michael Douglas for useful conversations.
The authors would like to thank
the Institute for Theoretical Physics and its staff for providing the
stimulating environment in which this work was begun.

\listrefs

\bye